\documentclass[manuscript]{acmart}
\AtBeginDocument{%
  \providecommand\BibTeX{{%
    \normalfont B\kern-0.5em{\scshape i\kern-0.25em b}\kern-0.8em\TeX}}}
\usepackage{graphicx,tabularx,subcaption}
\usepackage{booktabs}
\usepackage{multirow}
\usepackage[capitalise,nameinlink,compress]{cleveref}
\usepackage{xcolor}
\usepackage{comment}
\usepackage[english]{babel}
\usepackage{fontawesome5}
\usepackage{arydshln}
\usepackage{enumitem}
\graphicspath{{images/}}

\definecolor{main}{HTML}{aa72d4}    
\definecolor{sub}{HTML}{f2e8f8}     

\copyrightyear{2024} 
\acmYear{2024} 
\setcopyright{rightsretained} 
\acmConference[UMAP Adjunct '24]{Adjunct Proceedings of the 32nd ACM Conference on User Modeling, Adaptation and Personalization}{July 1--4, 2024}{Cagliari, Italy}
\acmBooktitle{Adjunct Proceedings of the 32nd ACM Conference on User Modeling, Adaptation and Personalization (UMAP Adjunct '24), July 1--4, 2024, Cagliari, Italy}\acmDOI{10.1145/3631700.3664886}
\acmISBN{979-8-4007-0466-6/24/07}

\begin{document}


\title[An Explanatory Model Steering System]{An Explanatory Model Steering System for  Collaboration between Domain Experts and AI}

\author{Aditya Bhattacharya}
\orcid{0000-0003-2740-039X}
\email{aditya.bhattacharya@kuleuven.be}
\affiliation{%
  \institution{KU Leuven}
  \city{Leuven}
  \country{Belgium}
}
\author{Simone Stumpf}
\orcid{0000-0001-6482-1973}
\email{Simone.Stumpf@glasgow.ac.uk}
\affiliation{%
  \institution{University of Glasgow}
  \city{Glasgow}
  \country{Scotland, UK}
}

\author{Katrien Verbert}
\orcid{0000-0001-6699-7710}
\email{katrien.verbert@kuleuven.be}
\affiliation{%
  \institution{KU Leuven}
  \city{Leuven}
  \country{Belgium}
}

\renewcommand{\shortauthors}{Bhattacharya, et al.}

\begin{abstract}
With the increasing adoption of Artificial Intelligence (AI) systems in high-stake domains, such as healthcare, effective collaboration between domain experts and AI is imperative. To facilitate effective collaboration between domain experts and AI systems, we introduce an Explanatory Model Steering system that allows domain experts to steer prediction models using their domain knowledge. The system includes an explanation dashboard that combines different types of data-centric and model-centric explanations and allows prediction models to be steered through manual and automated data configuration approaches. It allows domain experts to apply their prior knowledge for configuring the underlying training data and refining prediction models. Additionally, our model steering system has been evaluated for a healthcare-focused scenario with 174 healthcare experts through three extensive user studies. Our findings highlight the importance of involving domain experts during model steering, ultimately leading to improved human-AI collaboration.
\end{abstract}

\ccsdesc[500]{Human-centered computing~Human computer interaction (HCI)}
\ccsdesc[500]{Human-centered computing~Interaction design}
\ccsdesc[500]{Computing methodologies~Machine learning}

\keywords{Explainable AI, XAI, Interactive Machine Learning, IML, Explanatory Model Steering}


\begin{teaserfigure}
  \centering
  \includegraphics[width=0.6\linewidth]{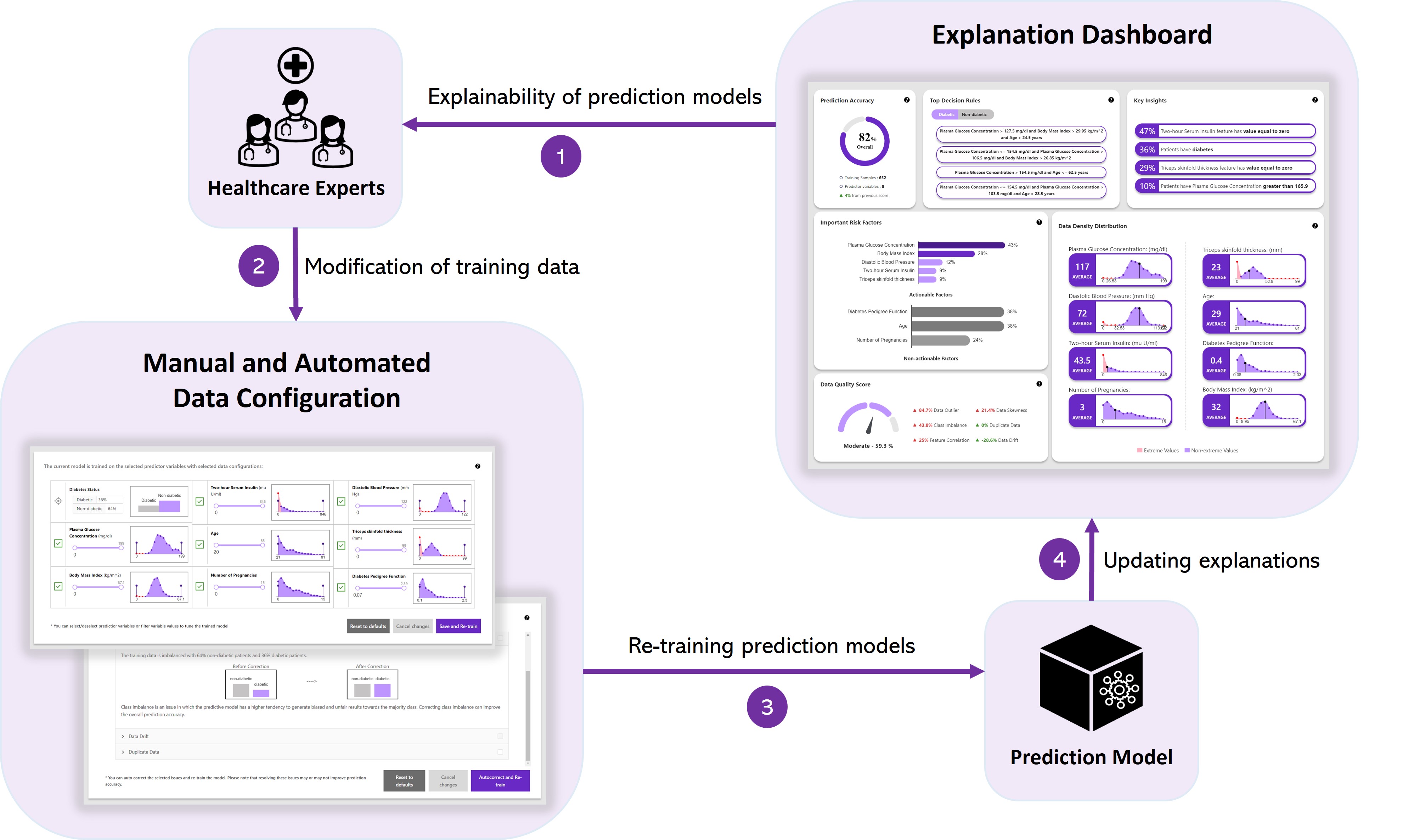}
  \caption{Our Explanatory Model Steering system consists of four aspects: (1) an explanation dashboard for providing explainability of prediction models, (2) domain expert driven model steering, (3) manual and automated data configuration approaches for fine-tuning the training data, and (4) updating prediction models and explanations on the configured data.  
  }
  \Description[Explanatory Model Steering for healthcare]{Our Explanatory Model Steering system consists of four aspects: (1) an explanation dashboard for providing explainability of prediction models, (2) domain expert driven model steering, (3) manual and automated data configuration approaches for fine-tuning the training data, and (4) updating prediction models and explanations on the configured data.}
  \label{fig:teasure_image}
\end{teaserfigure}

\maketitle




\section{Introduction}
The growing popularity of Artificial Intelligence (AI) systems in diverse application domains has led to an increased demand for including domain experts with little to no machine learning (ML) knowledge in the process of debugging and improving prediction models \cite{Rojo_2021, bhattacharya2024exmos, Bayar2021, kulesza_explanatory_2010, kulesza_principles_2015, teso_leveraging_2022}. Moreover, prior researchers have noted the advantages of a thorough domain knowledge possessed by domain experts for debugging and improving prediction models \cite{feuerriegel2020fair, Schramowski2020}. For instance, in healthcare, healthcare experts generally have a better understanding of patient's medical data, which is crucial for identifying biases and limitations in the training data used by prediction models~\cite{bhattacharya2024exmos, bhattacharya2023_technical_report}. Additionally, an effective collaboration with prediction models allow domain experts to better understand them and have higher confidence in their predictions \cite{lakkaraju2022rethinking}. 

To facilitate effective collaboration between domain experts and AI, we present an explanatory model steering system that relies on diverse Explainable AI (XAI) methods for explaining various aspects of ML systems~\cite{adadi2018peeking, BhattacharyaXAI2022, wang_designing_2019, pawar2020incorporating, Guidotti2018, anik_data-centric_2021, lakkaraju2022rethinking} and allows user-in-the-loop model steering for fine-tuning prediction models~\cite{Wang_KDD_2022, Dudley2018, teso_leveraging_2022, teso2019, Schramowski2020, bertrand_chi_2023}. The approach of explanatory steering provides a proficient human-centric solution to the challenges of obtaining rich feedback from domain experts for improving prediction models~\cite{teso_leveraging_2022}. More specifically, this approach facilitates domain experts in a better contextual understanding of the training data through interactive explanations. Unlike popularly adopted one-off explanations such as feature importance explanations \cite{BhattacharyaXAI2022, adadi2018peeking} or saliency maps \cite{simonyan2014deep, BhattacharyaXAI2022, adadi2018peeking}, multifaceted explanations (i.e., combination of different types of data-centric and model-centric explanations) \cite{bhattacharya2024exmos} included in our system, allow domain experts in identifying potentially misleading and biased predictors that can impact the prediction model.

Additionally, the system allows domain experts to steer prediction models by configuring the training data. It includes two distinct approaches for configuring the data: (1) manual configuration and (2) automated configuration. The manual configuration provides more control over the predictor variables. It allows domain experts to remove corrupt, biased, or unimportant predictor variables. With the manual configuration, domain experts can modify the upper and lower limits of predictor variables to filter anomalies present in the data. Meanwhile, the automated configuration requires less effort to configure the training data. It can identify data issues and offer potential corrections through automated algorithms.
Furthermore, our explanatory model steering system has been evaluated for a healthcare-focused scenario with 174 healthcare experts across three distinct user studies, each analysing different components of the system. The results of our user studies underscore the importance of combining different types of global explanations during model steering. Additionally, we found that domain experts with more domain experience are more confident and effective during model steering with manual configurations. This paper introduces our healthcare-focused model steering system, detailing its components and the technical architecture adopted in its development. 
A demonstration of this system can be accessed on YouTube. \footnote{Demo video is available on YouTube: \url{https://www.youtube.com/watch?v=-9itqTkyQ6s}}.

\section{Explanatory Model Steering System}
This section first discusses the two main components of our explanatory model steering system: (1) a multifaceted explanation dashboard that combines different types of explanation methods and (2) data configuration mechanisms that allow domain experts to steer prediction models by configuring the training data. Then, we present the technical implementation of the system. The source code of the system is open-sourced on GitHub: \anon[Code Repository]{\url{https://github.com/adib0073/EXMOS/}}.

\subsection{Explanation Dashboard}
Inspired by recent studies that demonstrated the effectiveness of multifaceted explanations achieved by combining different types of explanations~\cite{Bhattacharya2023} for enhancing user understandability~\cite{teso_leveraging_2022, teso2019, Schramowski2020, bertrand_chi_2023, kulesza_principles_2015, kulesza_2013, Wang_deepseer2023}, we designed and developed a explanation dashboard (\Cref{fig:explanation_dashboard}), which combined the following two explanation types:

\begin{figure*}[h]
\centering
\includegraphics[width=0.60\linewidth]{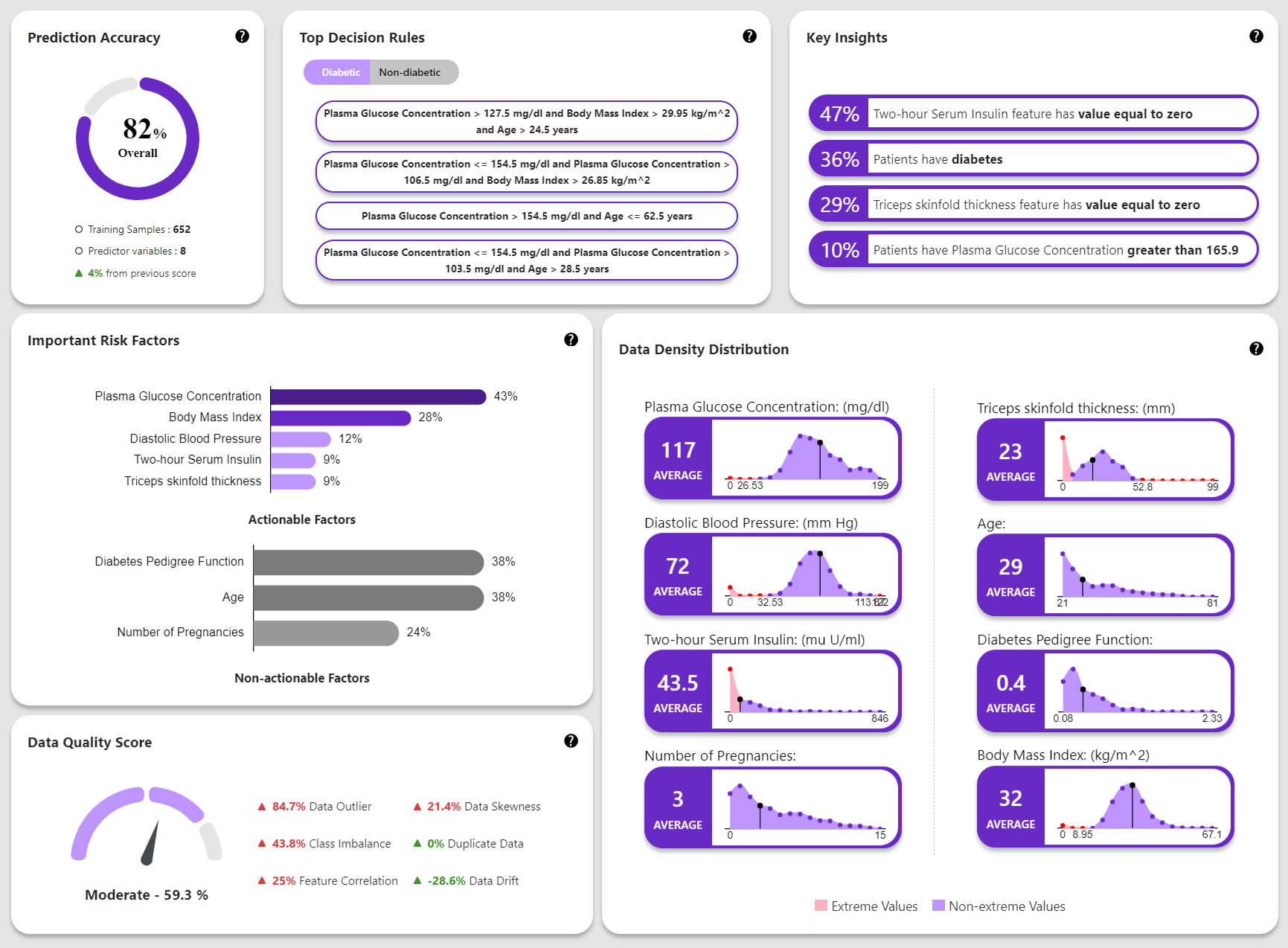}
\caption{Screenshot of the explanation dashboard which combines diverse data-centric and model-centric global explanations. The different visual components are designed based on the guidelines from Bhattacharya et al.~\cite{bhattacharya2024exmos}.}
\Description[Explanation Dashboard]{Design of multifaceted explanation dashboard from Bhattacharya et al. which combine data-centric and model-centric global explanations. The different visual components included in this design are: Top Decision Rules (TDR), Key Insights (KI), Important Risk Factors (IRF), Data Quality (DQ) and Data Density Distribution (DDD).}
\label{fig:explanation_dashboard}
\end{figure*}

\begin{itemize}[left=-0cm, noitemsep]
    \item \textit{Data-centric explanations}: These explanations elucidate different aspects of the training data to the user \cite{anik_data-centric_2021, BhattacharyaXAI2022}. We followed the design guidelines from Bhattacharya et al.~\cite{bhattacharya2024exmos} to include diverse data-centric global explanations that presented distinct patterns from the training data using descriptive statistics. It also presented the frequency distribution for each predictor variable used in the prediction model and the estimated quality of the data used in training the prediction model.
    \item \textit{Model-centric explanations}: Unlike data-centric explanations, model-centric explanations describe the various factors considered by the model for generating the predictions, such as feature weights or model hyper-parameters \cite{adadi2018peeking, BhattacharyaXAI2022}. We followed the design guidelines from Bhattacharya et al.~\cite{bhattacharya2024exmos} for including top decision rules generated by surrogate explainers~\cite{BhattacharyaXAI2022} and SHAP~\cite{lundberg2017unified} feature importances for obtaining the important actionable and non-actionable factors. 
\end{itemize}

Furthermore, the explanation dashboard also presents the model's overall prediction accuracy, along with the training data size and the number of predictor variables considered by the model. It also displays the change in prediction accuracy after the model is retrained after the data configurations. Domain experts are introduced to the explanation dashboard before the data configuration mechanism for effective model steering. 

\begin{figure*}[h]
\centering
\begin{subfigure}[b]{1.0\textwidth}
\centering
\includegraphics[width=0.62\linewidth]{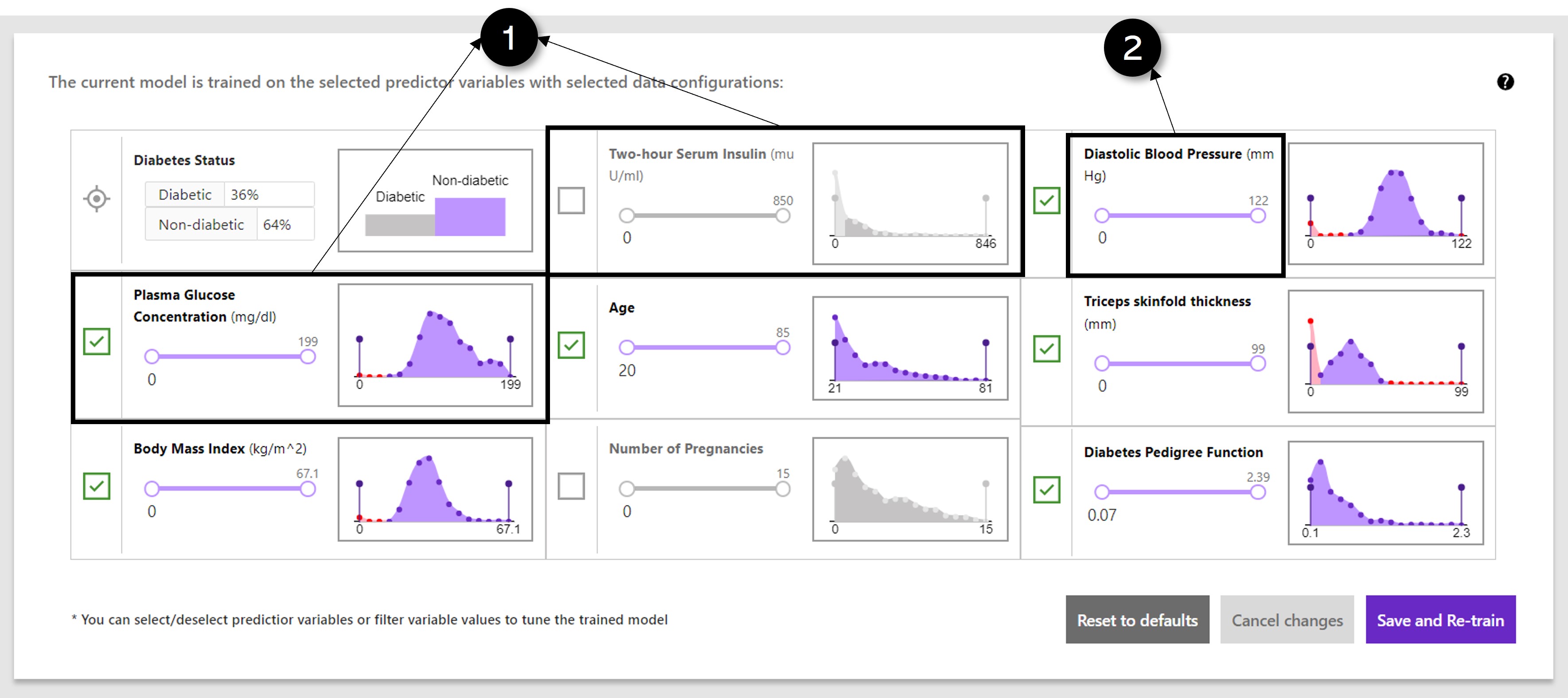}
\caption{Manual configuration design, which includes (1) feature selection control to include or exclude predictor variables and (2) feature filtering control to set the upper and lower limits for the predictor variables. }
\Description[Manual configuration design]{Manual configuration design, which included (1) feature selection control to include or exclude predictor variables and (2) feature filtering control to set the upper and lower limits for the predictor variables.}
\label{fig:manual_config}
\end{subfigure}
\par\bigskip
\begin{subfigure}[b]{1.0\textwidth}
\centering
\includegraphics[width=0.62\linewidth]{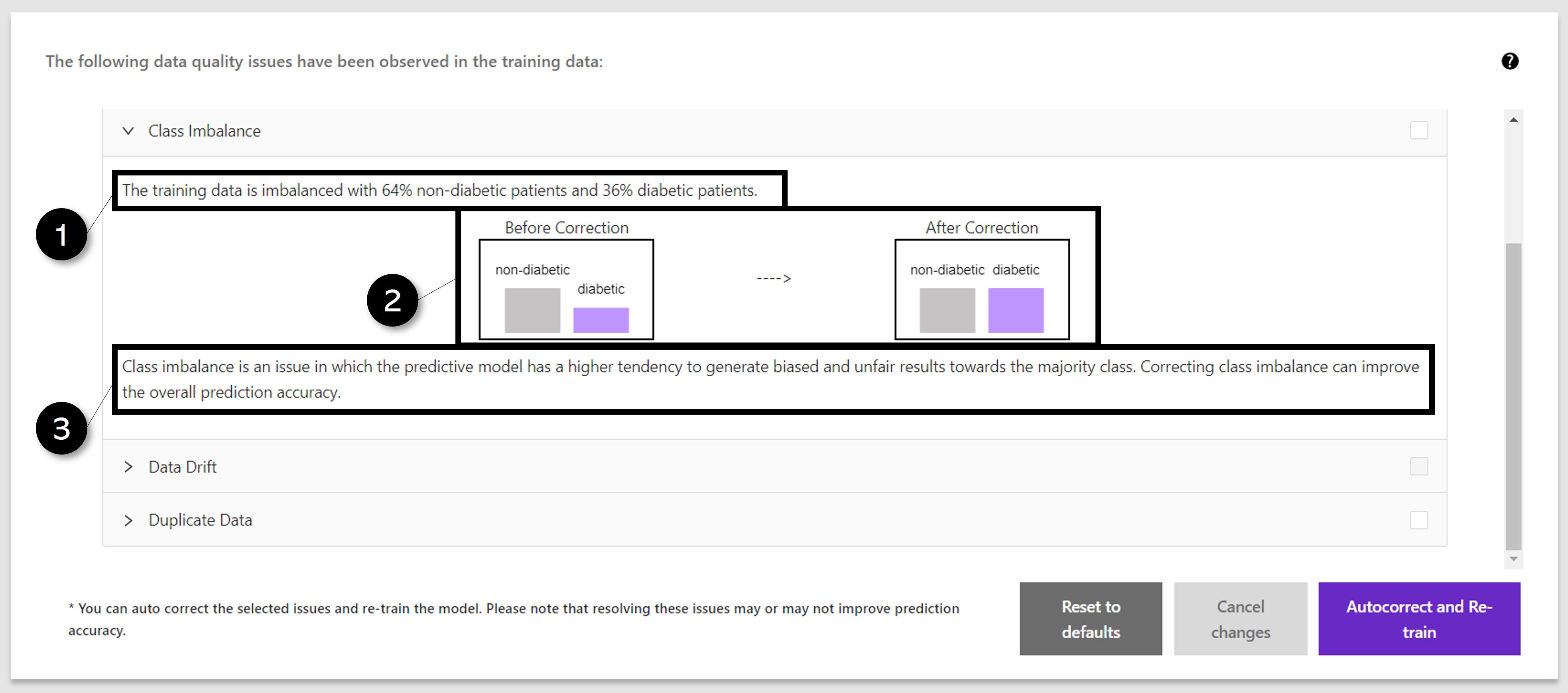}
\caption{Automated configuration design, which includes data issue explanations through (1) the quantified impact of these issues, (2) visualisations displaying before and after correction changes to the data or predictor variables and (3) description of the issue and how its correction can impact the model performance.}
\Description[Automated configuration design]{Automated configuration design, which included explanations of the data issues through (1) displaying the quantified impact of these issues, (2) visualisations displaying before and after correction changes to the data or predictor variables and (3) description of the issue and how its correction can impact the model performance.}
\label{fig:auto_config}
\end{subfigure}
\caption{Screenshots of the manual and automated configuration screens from the system. }
\Description[Manual and automated data configuration designs]{Data configuration designs for domain expert AI collaboration.}
\label{fig:dconfig}
\end{figure*}

\subsection{Data Configuration Approaches}
Our system provides domain experts with control over refining prediction models by leveraging their prior domain knowledge of the training data. It allows them to interact with the training data to identify and correct diverse data quality issues. Training data configuration can be achieved by the following distinct approaches:
\begin{enumerate}[left=-0cm, noitemsep]
    \item \textit{Manual configuration}: Following the principles of data-centric AI \cite{zha2023datacentric}, the manual configuration approach allows domain experts to perform feature selection, feature filtering and provide data guardrails to prevent abrupt changes to the training data. Using this mechanism, domain experts gain higher control over the training data for removing irrelevant features or potentially corrupt predictor variables. This approach also allows domain experts to identify the correct data ranges for the selected predictor variables for more reliable predictions.
    \item \textit{Automated configuration}: The automated configuration enables domain experts to select data issues that correction algorithms can automatically rectify \cite{zha2023datacentric}. It includes providing an explanation of the potential data issues and their impact on the prediction model. Then, to minimise the impact of these potential data issues, domain experts could select the potential issues for automated corrections with a single button click. However, despite requiring less effort than the manual mechanism, the automated method provides lesser control for configuring training data.
\end{enumerate}

\begin{figure*}[h]
\centering
\includegraphics[width=0.6\linewidth]{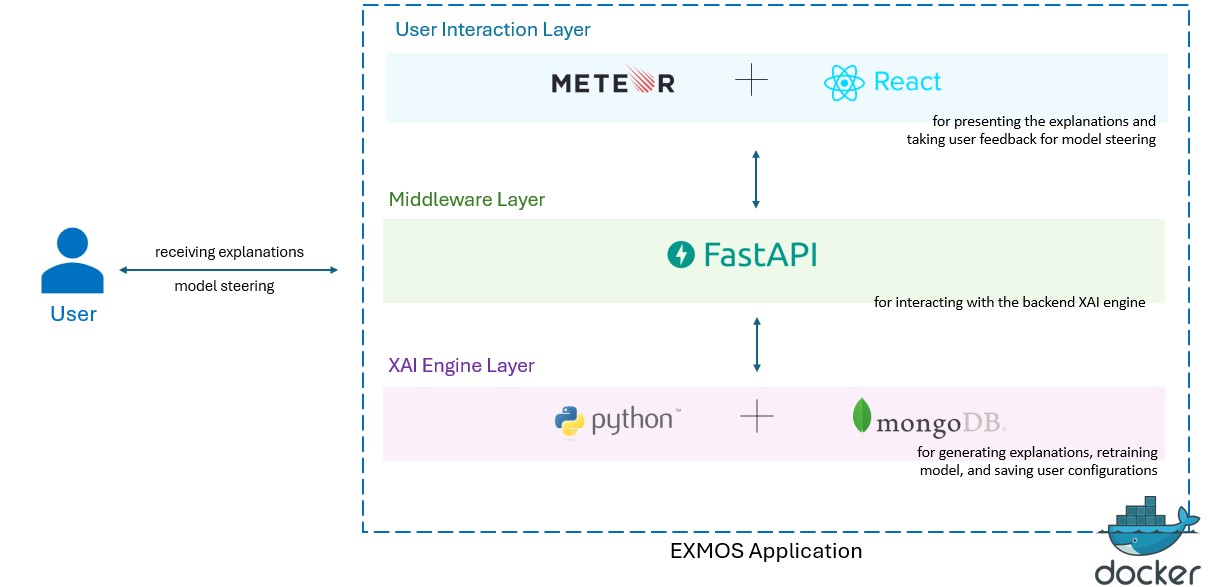}
\caption{The model steering application adopts a three-layered architecture to facilitate user interactions.}
\Description[Explanation Dashboard]{High-level technical solution architecture for model steering application.}
\label{fig:solution_diagram}
\end{figure*}

\subsection{Technical Implementation}
Our model steering application adopts a three-layered architecture, as illustrated in \Cref{fig:solution_diagram}, to facilitate user interactions:
\begin{enumerate}[left=-0cm, noitemsep]
    \item \textit{User Interaction Layer}: This layer utilizes Meteor's full-stack framework \cite{meteor} in conjunction with React.js 
    to provide interactive visual explanations to users. It also captures user feedback for model steering, enabling users to guide the explanation process.
    \item \textit{Middleware Layer}: Developed using the FastAPI framework \cite{fastapi}, this layer acts as a bridge between the user interface layer and the back-end XAI engine. It receives requests from the user interface, transmits them to the XAI engine for processing, and returns back the responses to the user interface layer.
    \item \textit{XAI Engine}: The XAI engine is the core of the application where the ML and XAI algorithms are applied. This layer is implemented in Python. It connects to a MongoDB 
    database to store training data and user configurations. This layer is responsible for refining the prediction model based on user feedback and regenerating the explanations based on the configured training data and the fine-tuned ML model. 
\end{enumerate}
To ensure scalability and reproducibility, the entire application is containerised using Docker and deployed on our \anon{KU Leuven} managed servers.

\section{Use Case: Steering Diabetes Prediction System}
We instantiated the generic design of our model steering system for a healthcare-focused diabetes prediction system. In this scenario, healthcare experts were provided with a model steering system that explained a diabetes prediction model using the Pima Indians dataset \cite{Smith1988-rv}. Through our experiments, we analysed how healthcare experts could steer the prediction model using this system. We conducted three user studies involving 174 healthcare experts in total to analyse the effectiveness of the explanation dashboard and the data configuration approaches during model steering. The findings from our extensive user studies highlight the importance of the multifaceted explanations and the various data configuration approaches for effective collaboration between domain experts and AI. 
Further details of the first two user studies and their in-depth findings have been published in ACM CHI 2024 \cite{bhattacharya2024exmos}. The detailed description of the third user study and its key findings is currently undergoing review for publication in another ACM conference paper.  In general, the model steering system facilitates the involvement of domain experts during model steering,
ultimately leading to improved human-AI collaboration.


\section{Conclusion}
In this paper, we introduce an Explanatory Model Steering system that facilitates effective collaboration between domain experts and AI. This system allows domain experts to steer prediction models using manual and automated data configuration approaches. It involves domain knowledge driven configuration of the training data by domain experts for fine-tuning prediction models. While evaluated for the healthcare domain, the applicability of such a system can be extended to other application domains.

\begin{acks}
We thank Maxwell Szymanski and Robin De Croon for their valuable feedback on this research. This research was supported by Research Foundation–Flanders (FWO grants G0A4923N and G067721N) and KU Leuven Internal Funds (grant
C14/21/072)~\cite{BhattacharyaCHIDC}. We also thank the participants of our user studies for their contribution.
\end{acks}

\bibliographystyle{ACM-Reference-Format}
\bibliography{references}


\end{document}